\documentclass[preprint,a4paper,showpacs,aps,amsmath,amssymb]{revtex4}
\usepackage{graphicx}
\usepackage{amsmath}
\usepackage{amssymb}

\begin{document}

\title{A semiclassical trace formula for the canonical partition function
of one dimensional systems}

\author{Fernando Parisio and M. A. M. de Aguiar}

\affiliation{Instituto de F\'{\i}sica `Gleb Watghin', Universidade
Estadual de Campinas, Caixa Postal 6165, 13083-970, Campinas,
S\~ao Paulo, Brazil}

\begin{abstract}
We present a semiclassical trace formula for the canonical partition
function of arbitrary one-dimensional systems. The approximation
is obtained via the stationary exponent method applied to the
phase-space integration of the density operator in the coherent
state representation. The formalism is valid in the low temperature
limit, presenting accurate results in this regime. As illustrations
we consider a quartic Hamiltonian that cannot be split into kinetic
and potential parts, and a system with two local minima. Applications 
to spin systems are also presented.

\pacs{03.65.Sq,05.30.-d }
\end{abstract}
\maketitle
\section{Introduction}

The quantum partition function ${\cal Z}$ is the basic physical
quantity from which all ensemble averages are derived in the
canonical formalism \cite{salinas,garrod,mandl}. It is given by the
trace of the density operator ${\cal Z}={\rm Tr}\,[\hat{\rho}]$,
that can be expressed either as a discrete sum in the microstate
energies ${\cal Z}=\sum_n e^{-\beta E_n}$ or as a continuum
integral, ${\cal Z}=1/h\int{\rm d}q\,{\rm d}p \,\rho$, where $\rho$
is some representation of the density operator in phase-space, like
the Wigner function \cite{wigner} or the coherent state average of
$\hat{\rho}$. In classical mechanics the density is simply
$\rho_{class} = e^{-\beta {\cal H}}$, where ${\cal H}$ is the
classical Hamiltonian.

Because of the difficulties in the exact evaluation of ${\cal Z}$, a
variety of formal developments have appeared in the literature to
include quantum effects in statistical mechanics, without exactly
solving the Schrodinger equation. Semiclassical methods, in
particular, have attracted a lot o attention
\cite{feynman,miller3,joras1,joras2,miller1,miller2}. From a more
basic point of view, it is well-known that classical statistical
mechanics is, in most of cases, unable to reproduce the experimental
behavior for $T \rightarrow 0$. A paradigmatic example is the failure
at low temperatures of Dulong-Petit law for the specific heat of
solids, or more generally, the failure of classical systems in obeying
the third law of thermodynamics \cite{salinas,garrod,mandl}. Roughly
speaking, this is due to quantum discreteness that becomes relevant in
this regime. Given this strong discrepancy between classical and
quantum statistical mechanics for $T \rightarrow 0$, it is reasonable
to ask what happens in a semiclassical formulation, since it should be
somewhere in between the two theories. Therefore, it is desirable to
obtain an explicit semiclassical expression for the canonical
partition function, from which thermodynamic quantities such as
internal energy and entropy can be derived in a closed analytical form
in terms of classical ingredients.

A straight, but not particularly effective, procedure in this
direction is to replace the exact quantum levels by semiclassical ones
in the discrete summation. For 1-D systems, expressions for the energy
levels in the WKB approximation are relatively easy to obtain, but the
corresponding partition function is inaccurate at low
temperatures. More elaborated examples are the Wigner-Kirkwood method,
which consists of an expansion of the density matrix in powers of
$\hslash \beta=\hslash/k_BT$ \cite{kirk}; approximate path integral
representations \cite{feynman}; and phase-space sampling with
semiclassical quasi-probability distributions
\cite{miller3,joras1,joras2}.  All these methods use classical
ingredients to represent the density operator, but the procedure
presented in \cite{miller3} is perhaps the most direct, in the sense
that it simply replaces the classical Boltzmann weight by a
semiclassical one. In this paper we shall consider a similar
method. The procedure is based on the recognition that the evolution
operator in quantum mechanics and the Boltzmann operator in
statistical physics are formally similar. Explicitly we have the
correspondence $e^{ -i\hat{{\cal H}} t /\hslash}\rightarrow e^{-\beta
\hat{{\cal H}} }$ for $t \rightarrow -i \tau$, with $\tau \equiv
\hslash \beta$ (the so-called Wick rotation). The density operator
$\hat{\rho}$ is obtained when such a rotation is applied to the
diagonal quantum propagator and the partition function is written as a
phase-space (or configuration space \cite{miller1,miller2}) integral
that can be calculated numerically. The ensemble averages are,
consequently, also expressed in the form of integrations (see
equations (2.16) to (2.18) in \cite{miller3}).  This method, in spite
of its usefulness in numerical calculations, does not allow for a
detailed analysis of the functional dependence of thermodynamic
potentials, and other related physical quantities of interest, with
the temperature. In this work we show that, in the low temperature
limit, the phase space integrals appearing in the semiclassical
expressions can actually be performed analytically. The resulting
representation of the partition function in terms of classical
quantities is similar to a trace formula, involving only equilibrium
points and periodic orbits. In order to obtain the semiclassical
partition function we shall use the coherent state
representation. This choice enables the formalism to be applied in
more general situations involving internal degrees of freedom such as
spins. 

We believe that such a direct semiclassical formula should be of use
in theoretical studies and also in numerical applications. Besides,
some traditional methods like the Wigner-Kirkwood expansion converge
very slowly for small $T$, justifying the particular interest in this
region of temperatures.  As we shall see, however, the semiclassical
calculation of ${\rm Tr}[\hat{\rho}]$ naturally leads to the
complicated problem of summing over all periodic orbits of a related
classical system. This is a common difficulty in practical
applications of semiclassical trace formulas. Fortunatelly, due to the
effects of the Wick rotation in the classical equations of motion, we
shall see that this problem is not present in some important
situations, e. g., one-dimensional potentials with a single minimum.

The paper is organized as follows: in section II we review the
concepts of Wick rotation, coherent states and semiclassical
propagator. Next we derive the semiclassical formula for the partition
function and give explicit expressions for potentials with a single
well and multiple wells. In section IV we apply these formulas for
three specific systems and compare our results with exact quantum and
classical calculations and with other approximations. In section V we
outline the extension of the formalism to spin systems. Finally,
section VI is devoted to some concluding remarks.

\section{Preliminary Definitions}

The partition function can be written in terms of the quantum density
of particles in the coherent state representation as
\begin{equation}
{\cal Z}= \int\frac{{\rm d}q\,{\rm d}p}{2\pi \hslash}\, \langle
z|\,e^{ -\beta \hat{\cal H} }\,| z \rangle = \int\frac{{\rm d}q\,{\rm
d}p}{2\pi \hslash}\,\rho(q,p,\beta)\;.
\end{equation}
\label{Zpq1}
Our procedure to calculate the semiclassical limit of ${\cal Z}$
consists of two main steps: first we perform a Wick rotation in the
quantum propagator $\langle z|\,e^{ -i \hat{{\cal H}}t / \hslash
}\,| z \rangle$. Next we replace the propagator by its semiclassical
expression and calculate the integral over $q$ and $p$ explicitly by
the saddle point method to obtain ${\cal Z}_{sc}$. In the following
subsections we review the basic ingredients of this approach: the
Wick rotation, the coherent state representation and the
semiclassical propagator. In the next section we calculate ${\cal
Z}_{sc}$.

\subsection{Wick rotation and classical dynamics}

Semiclassical formulas contain classical quantities, such as actions
and their derivatives, thus our first step is to understand the
effect of the rotation $t \longrightarrow -i \tau$ on these classical
ingredients. We start with Newton's second law. Consider the pair of
transformations
\begin{equation}
t = -i \tau\; ,\;\;  V(q) =
-\overline{V}(q) \;,
\label{newton}
\end{equation}
leading to
\begin{equation}
\frac{{\rm d}^2q}{{\rm d}t^2}=-\frac{{\rm d} V}{{\rm d} q} \;
\longrightarrow \; \ddot{q}=\frac{{\rm d} V}{{\rm d} q} \;
\longrightarrow \; \ddot{q}=-\frac{{\rm d} \overline{V}}{{\rm d}
q}\; ,
\end{equation}
where dot means derivative with respect to $\tau$. We see that in
order to get back the usual form of the equation after the Wick
rotation we have set $V = -\overline{V}$, which means that the
classical dynamics with complex time is formally equivalent to
real time dynamics with the reversed potential $\overline{V}$
\cite{feynman,miller1}. The effect on the classical action is as
follows ($m=1$)
\begin{equation}
S=\int_0^t {\rm d}t'{\cal L}=\int_0^t {\rm d}t'\left[
\frac{1}{2}\left( \frac{{\rm d} q}{{\rm d} t} \right)^2 - V\right] \rightarrow
i\int_0^{\hslash \beta} {\rm d}\tau'\left( \frac{1}{2}\dot{q}^2 -
\overline{V}\right) =i\int_0^{\hslash \beta} {\rm d}\tau'\,
\overline{{\cal L}}=i\, \overline{S} \;,
\end{equation}
where it is understood that $\overline{\cal L}$ and $\overline{S}$ in the
right-hand side are evaluated for a final time $\hslash \beta$ and
with the reversed potential $\overline{V}$. There is, however, a
more general approach, that we may call `canonical Wick rotation',
which is constructed in phase-space and does not rely on the
existence of the function $V(q)$. Starting from Hamilton's
equations and making
\begin{equation}
t =-i \tau \; ,\;\; {\cal H} = iH
\label{hamiltons}
\end{equation}
we obtain
\begin{equation}
\frac{{\rm d} q}{{\rm d} t}=\frac{\partial {\cal H}}{\partial p}\; ,
\;\; \frac{{\rm d} p}{{\rm d} t}=-\frac{\partial {\cal H}}{\partial
q} \;\; \rightarrow \;\; \dot{q}=-i\frac{\partial {\cal H}}{\partial
p}\;,\;\; \dot{p}=i\frac{\partial {\cal H}}{\partial q} \;\;
\rightarrow \;\; \dot{q}=\frac{\partial H}{\partial p}\;,\;\;
\dot{p}=-\frac{\partial H}{\partial q}\;,
\end{equation}
which preserve the original form of the canonical equations. If the
Hamiltonian can be written in `Euclidean' form ($p^2/2+V$) the
operations (\ref{newton}) and (\ref{hamiltons}) produce the same
overall result. However, for more general Hamiltonians, for example
with force fields not derivable from a potential or involving spin
degrees of freedom, only the last pair of transformations is
applicable. The classical action is invariant under these
transformations,
\begin{equation}
S=\int_0^t {\rm d}t'\left[ p\left( \frac{{\rm d} q}{{\rm d}
t} \right)-{\cal H}\right] \rightarrow
\int_0^{\hslash \beta} {\rm d}\tau'\left( p\dot{q}-H \right) =
S\;.
\end{equation}
In deriving our semiclassical partition function we shall employ the
canonical Wick rotation because of its generality. As illustrations
we apply the transformation (\ref{hamiltons}) to a non-Euclidean
Hamiltonian and the transformation (\ref{newton}) to a system with a
well defined potential in section IV.

\subsection{Coherent state representation}

Here we briefly describe some properties of the coherent state (or
Bargman) representation of quantum mechanics. Canonical coherent
states $\{|z\rangle \}$ can be expressed as an infinite sum of
harmonic oscillator number states $\{|n\rangle \}$:
\begin{equation}
|z\rangle =e^{-|z|^2/2}\sum_{n=0}^\infty \frac{z^n}{\sqrt{n!}}\,
|n\rangle \;, \;\; \mbox{where} \;\; z =  \frac{1}{\sqrt{2}}\left(
\frac{q}{b}+i \,\frac{bp}{\hbar} \right) \; . \label{z}
\end{equation}
The real numbers $q$ and $p$ are the expectation values of the
corresponding quantum operators $\hat{q}$ and $\hat{p}$ in the
state $|z\rangle$, while the parameter $b$ is proportional to the
uncertainty in position, $b=\sqrt{2} \, \Delta q$, and is related
to the frequency of the associated harmonic potential by
$b^2=\hslash/m\omega$. The resolution of unity can be written in
terms of the set $\{|z\rangle \}$ as
\begin{equation}
\hat{I}=\int\frac{{\rm d}\,q{\rm d}p}{2\pi \hslash}\; |z \rangle
\langle z| \equiv \int\frac{{\rm d}\,z^*{\rm d}z}{2\pi i}\; |z
\rangle \langle z|\;.
\end{equation}
In the coherent state representation the density of particles in
phase-space reads
\begin{equation}
\rho(q,p,\beta)=\langle z | e^{-\beta \hat{{\cal H}}} |z
\rangle\;,
\end{equation}
whose corresponding quantity under the inverse transformation
$\tau=i t$ is the diagonal form of the coherent state propagator
$K(q,p,t)=\langle z | e^{-i\hat{{\cal H}}t/\hslash} |z \rangle$.

We finish this subsection by defining the canonically conjugated
variables $u(t)=[q(t)/b+i \,bp(t)/\hbar]/\sqrt{2}$ and
$v(t)=[q(t)/b-i\,bp(t)/\hbar]/\sqrt{2}$ in terms of which
Hamilton's equations read
\begin{equation}
\frac{{\rm d} v}{{\rm d} t}=\frac{i}{\hslash}\frac{\partial {\cal
H}}{\partial u}\;, \;\; \frac{{\rm d} u}{{\rm d}
t}=-\frac{i}{\hslash}\frac{\partial {\cal H}}{\partial v}\;.
\label{hamilton}
\end{equation}
These variables are particularly convenient to write down the
semiclassical limit of the coherent state propagator.

\subsection{Semiclassical coherent state propagator}

The set of coherent states forms a non-orthogonal over-complete
basis, since each state in the set can be written as a linear
combination of the others. This over-completeness implies the
existence of several forms of the path integral formulation for the
propagator, all equivalent quantum mechanically, but each leading to
a slightly different semiclassical limit. Klauder and Skagerstam
\cite{Klau85} proposed two basic forms for the coherent state path
integral. The semiclassical limit of these propagators
were considered in \cite{marcus} where it was shown that both
propagators can be written in terms of classical complex
trajectories, each governed by a different classical representation
of the Hamiltonian operator $\hat{{\cal H}}$: the P representation
in one case and the Q representation in other. The phase appearing
in these semiclassical formulas turns is not just the action
of the corresponding complex classical trajectory, but it also
contains a `correction term' that comes with different signs in each
formula. In \cite{marcus} it was also suggested that a semiclassical
representation involving directly the Weyl representation of
$\hat{{\cal H}}$, or the {\it classical Hamiltonian} ${\cal H}$,
could probably be constructed, and a formula for this representation
was conjectured. This conjecture, along with the corresponding
quantum mechanical path integral representation, has been recently
proved \cite{coelho} using the translation and reflection operators
studied in \cite{ozorio}. In this paper we shall adopt this later
semiclassical expression, since it is the simplest (although not
always the most accurate \cite{Pol03}) of the three known formulas.
This semiclassical expression for the propagator connecting an
initial state $|z '\rangle$ to a final state $|z ''\rangle$ is given
by
\begin{eqnarray}
\langle z''|e^{ -\frac{i}{\hslash} \hat{\cal H} t }| z'
\rangle_{sc}=\sqrt{\frac{1}{M_{vv}}}\;
\exp\left\{\frac{i}{\hbar}\Phi(v'', u',
t)-\frac{1}{2}(|u'|^2+|v''|^2) \right\}, \label{zzprop}
\end{eqnarray}
where there is an implicit sum over the, usually complex,
trajectories satisfying Hamilton's equations (\ref{hamilton}) with
boundary conditions $u'\equiv u(0)=z'$, $v'' \equiv v(t)=z''^*$. The
Hamiltonian entering in formula (\ref{zzprop}) is the function
${\cal H}$ describing the corresponding classical system, and the
function appearing in the exponent is the complex action given by
\begin{equation}
\Phi(v'', u', t)=\int_{0}^{t}\left[\frac{i\hbar}{2}\left(v
\frac{{\rm d} u}{{\rm d} t}-u\frac{{\rm d} v}{{\rm d}
t}\right)-{\cal H} \right]{\rm d}t'-\frac{i\hbar}{2}
\left[u'v'+v''u''\right]\; , \label{action1}
\end{equation}
where we have set $u''\equiv u(t)$ and $v' \equiv v(0)$. Note that,
due to the complex character of the classical orbits, in general, we
have $v' \ne z'^*$ and $u'' \ne z''$. Finally, the pre-factor in
formula (\ref{zzprop}) is written in terms of the tangent matrix,
whose elements are given by the following relation
\begin{align}
   \label{m}
        \begin{pmatrix}
           \delta u''  \\ \\
           \delta v''
        \end{pmatrix}
    \equiv  \begin{pmatrix}
              M_{uu} & M_{uv} \\ \\
              M_{vu}  & M_{vv}
            \end{pmatrix}
            \begin{pmatrix}
               \delta u' \\ \\
               \delta v'
            \end{pmatrix}
\;,
\end{align}
where $(\delta u'$, $\delta v')$ denote small initial deviations
from the classical orbit and $(\delta u''$, $\delta v'')$ the
corresponding deviations after a propagation time $t$. It is
simple to show that $\det[M]=1$ and that the following relations
are valid
\begin{equation}
\frac{i \hslash }{M_{vv}}=-\partial^2 \Phi/\partial u'\partial v''
\; , \;\; \frac{1}{M_{uv}}=\frac{\partial^2 \Phi/\partial
u'\partial v''}{\partial^2 \Phi/\partial {v''}^2}\; ,\;\;
\frac{1}{M_{vu}}=-\frac{\partial^2 \Phi/\partial u'\partial
v''}{\partial^2 \Phi/\partial {u'}^2}\;. \label{mdphi}
\end{equation}
%

\section{The semiclassical canonical partition function}

We proceed to make the transformations $t = -i\tau$ and ${\cal H}
= iH$ in Eq.(\ref{zzprop}) to get the semiclassical density of
particles in phase-space. Since the complex action and,
consequently, the tangent matrix elements remain invariant, the
semiclassical density of particles at a given temperature is
\begin{eqnarray}
\rho_{sc}=\langle z'|e^{ -\beta \hat{\cal H} }| z'
\rangle_{sc}=\sqrt{\frac{1}{M_{vv}}}\;\exp\left\{\frac{i}{\hbar}\Phi(v'',
u', \tau)-z'^*z' \right\}\; . \label{zzden}
\end{eqnarray}
We remark that Hamilton's equations (\ref{hamilton}) become
\begin{equation}
\dot{v}=\frac{i}{\hslash}\frac{\partial H}{\partial u}\;, \;\;
\dot{u}=-\frac{i}{\hslash}\frac{\partial H}{\partial v}\; ,
\label{hamilton3}
\end{equation}
with the boundary conditions (for the diagonal propagator)
\begin{equation}
 u(0)=u'=z'\; , \;\;v(\tau)=v''=v(\hslash \beta)=z'^*\;.
\label{cc}
\end{equation}
The semiclassical partition function is given by
\begin{equation}
{\cal Z}_{sc}= \int \frac{du'\,dv''}{2\pi
i}\sqrt{\frac{1}{M_{vv}}}\;\exp\left\{\frac{i}{\hbar}\Phi(v'', u',
\tau)_{v''=z'^* ,u'=z' }\;-z'^*z' \right\}\; . \label{int}
\end{equation}
In order to evaluate this integral by stationary exponent
approximation we shall assume the low temperature limit $\beta
>> 1$. The stationarity conditions are
\begin{equation}
\frac{i}{\hslash}\frac{\partial \Phi}{\partial
u'}-z'^*=v'-z'^*=0\; ,\;\; \frac{i}{\hslash}\frac{\partial
\Phi}{\partial v''}-z'=u''-z'=0
 \;  \label{per1}
\end{equation}
where we used the relations $\partial \Phi /\partial u'=-i\hbar
v'$ and $\partial \Phi /\partial v''=-i\hbar u''$ \cite{marcus}.
We get
\begin{equation}
v'=z'^* \; ,\;\; u''=z' \; . \label{per2}
\end{equation}
Therefore, the trajectories must satisfy the four
conditions (\ref{cc}) and (\ref{per2}). It can be shown that these
restrictions demand the contributing orbits to be {\it real and
periodic}. Next we expand the exponent in (\ref{int}) around the
stationary trajectories, labeled with the subscript `$0$', up to
second order and evaluate the pre-factor on the contributing
solutions. We obtain
\begin{equation}
\begin{array}{ll}
{\cal Z}_{sc}&= \displaystyle{
\left(\frac{e^{\frac{i}{\hslash}\Phi-u'v''}}{\sqrt{M_{vv}}}\right)_0
\int
\frac{{\rm d}(\Delta u')\,{\rm d}(\Delta v'')}{2\pi i} }\\ \\
&\times \exp{\left \{ \frac{i}{2\hslash}\left( \frac{\partial^2
\Phi}{\partial u'^2}\right)_0 \Delta u'^2+\frac{i}{2\hslash}\left(
\frac{\partial^2 \Phi}{\partial v''^2}\right)_0 \Delta v''^2
+\left(\frac{i}{\hslash} \frac{\partial^2 \Phi}{\partial u'
\partial v'' }-1\right)_0 \Delta u' \Delta v'' \right \}}\; ,
\end{array}
\end{equation}
where $\Delta u'=u'-u'_0$ and $\Delta v''=v''-v''_0$. The Gaussian
integration gives
\begin{equation}
{\cal
Z}_{sc}=\left(\frac{e^{\frac{i}{\hslash}\Phi-u'v''}}{\sqrt{M_{vv}}}\right)_0
\left[ \left(\frac{i}{\hslash} \frac{\partial^2 \Phi}{\partial u'
\partial v'' }-1\right)^2+\frac{1}{\hslash^2} \frac{\partial^2
\Phi}{\partial u'^2} \frac{\partial^2 \Phi}{\partial v''^2}
\right]_0^{-1/2} \; .
\end{equation}
From relations (\ref{mdphi}) we obtain the following
compact expression for the semiclassical partition function
\begin{equation}
{\cal Z}_{sc}=\sum_j\frac{e^{\frac{i}{\hslash}
\Phi_0^{(j)}-{u'}^{(j)}_0{v''}^{(j)}_0}}{\sqrt{{\rm
Tr}[M^{(j)}_0]-2}}\; , \label{Zsc1}
\end{equation}
where the sum runs over the real periodic trajectories with period
$\hbar\beta$ and Tr$[M]=M_{uu}+M_{vv}$. Note that, since the
contributing trajectories are periodic, the second term in the
right-hand side of expression (\ref{action1}) is exactly $-i\hslash
u'v''$, so we can write the exponent in equation (\ref{Zsc1}) as
\begin{equation}
\frac{i}{\hslash}
\int_{0}^{\tau}\left[\frac{i\hbar}{2}\left(v
\dot{u}-u\dot{v}\right)-{\cal H} \right]{\rm d}\tau'=\frac{i}{\hslash}\int_{0}^{\tau}
(p\dot{q}-{\cal H}){\rm d}\tau'-\frac{i}{2\hslash}pq|_{0}^{\tau}= \frac{i}{\hslash}S\;
\end{equation}
where we have used the periodicity in the last equality. Finally, we
can write the expression for the semiclassical partition function as
\begin{equation}
{\cal Z}_{sc}={\rm Tr}[\hat{\rho}_{sc}]=\sum_j
Z_{sc}^{(j)}=\sum_j\frac{e^{\frac{i}{\hslash} S_0^{(j)}}}{\sqrt{{\rm
Tr}[M^{(j)}_0]-2}}\; . \label{Zsc}
\end{equation}
We remark that, as is usual in stationary phase evaluations, there
may exist divergent contributions corresponding to spurious
stationary orbits. These orbits must not be taken into account in
the semiclassical formula since their inclusion through the
deformation of the original integration contours in (\ref{int})
could not be justified by Cauchy's Integral Theorem. Note that the
expected divergence in the infinite temperature limit occurs for
$\tau \rightarrow 0$, when $M$ becomes the identity matrix and
Tr$[M^{(j)}_0] \rightarrow 2$. Finally, we recall that, had we used
transformation (\ref{newton}) instead of (\ref{hamiltons}) the above
formula would have $-\overline{S}/\hslash$ as the argument of the
exponential. This alternative form is used in the next subsections.

\subsection{Single Well Potentials}

Now let us assume that there is a potential function $V$ which has a
single minimum (corresponding to a maximum of $\overline{V}$). Due
to the inversion in concavity, there are no periodic trajectories,
except for the trivial orbit that, without loss of generality, we
place at $u = v= 0$ (or $q = p = 0$). So, in this important case,
the summation in (\ref{Zsc}) reduces to a single term calculated on
the trivial orbit:
\begin{equation}
{\cal Z}_{sc}=\frac{e^{-\frac{1}{\hslash}\overline{S}_0}}{\sqrt{{\rm
Tr}[M_0]-2}}\; . \label{Zsc0}
\end{equation}
Formula (\ref{Zsc0}) can be evaluated more explicitly.
We have the following relations for the time evolution of
small perturbations:
\begin{equation}
\delta \dot{q}=\delta p\;,\;\; \delta \dot{p}=-\frac{\partial^2 \overline{V}}
{\partial q^2}\delta q\;.
\end{equation}
For a non-trivial trajectory the coefficient $\partial^2 \overline{V}/
\partial q^2 \equiv \overline{V}''$ is a time dependent function, since 
it must be evaluated on the trajectory $q(\tau)$, but for the trivial 
orbit it is simply a constant calculated at the maximum of $\overline{V}$,
and the above equations can be readily integrated. The result is
${\rm Tr}[M_0]=2\cos(\sqrt{ \overline{V}''}\;\tau)= 2\cosh(\sqrt{
V''}\;\tau)$, where  $ V''>0$. This leads to the following
semiclassical partition function
\begin{equation}
{\cal Z}_{sc}=\frac{e^{-\beta V(0)}}{2\sinh\left(\frac{1}{2} \hslash
\sqrt{ V''} \beta \right)}\;.
\end{equation}
The above expression is simply the partition function for the
harmonic oscillator that best fits $V$ at $q=0$. This result turns
out to be too poor and the conclusion is that a higher order
approximation is needed in this case. In section IV we illustrate
such a higher order expansion for a specific example. In contrast,
we shall show that the second order semiclassical partition function
given by (\ref{Zsc}) already gives non-trivial results in the case
of potentials with multiple minima.


\subsection{Multiple Well Potentials}
Let us assume that the potential $V$ has $N$ minima, located at
$q_\sigma$, with $\sigma=1,2,\dots,N$, and $N-1$ local maxima. In
this case there are non-trivial periodic orbits between two
successive maxima of $\overline{V}$. In addition each equilibrium
point corresponds to a trivial orbit that, in principle, should
contribute to the evaluation of ${\cal Z}_{sc}$. However, the $N-1$
stable equilibrium points of $\overline{V}$ give rise to terms
proportional to $1/\sin\left(\frac{\hslash}{2} \sqrt{\overline{V}''}
\beta \right)$, $\overline{V}'' > 0$, which have an oscillatory and
divergent behavior for $\beta \rightarrow \infty$, and correspond to
the spurious saddle points we have mentioned. Discarding these
contributions we get
\begin{equation}
{\cal Z}_{sc}=\sum_{\sigma = 1}^{N}\frac{e^{-\beta
V(q_\sigma)}}{2\sinh\left(\frac{1}{2} \hslash \sqrt{ V''|_\sigma}
\beta \right)}+{\sum_{\mu}}'\frac{e^{-\frac{1}{\hslash}
\overline{S}_0^{(\mu)}}}{\sqrt{{\rm Tr}[M^{(\mu)}_0]-2}}\;,
\label{Zmm}
\end{equation}
where we have split the sum in the trivial unstable equilibria (with
respect to $\overline{V}$) and the non-trivial periodic orbits
(denoted by the primed sum). We call the first term the harmonic
contribution and the second term the tunneling contribution, since
the $\mu$-orbits connect successive minima in the original
potential. The calculation of the tunneling terms requires a careful
procedure. For a fixed temperature $T$ we must seek, in each well of
$\overline{V}$, orbits with period $\tau=\hslash/k_BT$ and sum their
contributions. Note also that these orbits should, in principle,
contribute to the partition function at temperatures $T/2$,
$T/3$,..., $T/n$, corresponding to the propagation times $2\tau$,
$3\tau$,..., $n\tau$. However, these multiple traversal period
orbits have larger and larger actions for increasing values of $n$,
causing their contribution to fall off exponentially.

\subsection{Connection with thermodynamics}

The connection between the canonical ensemble in statistical
mechanics and thermodynamics can be made through relation ${\cal
Z}=\exp(-\beta f)$, where $f$ is the Helmholtz free energy per
particle, whose semiclassical expression reads
$f_{sc}=-k_BT\ln({\cal Z}_{sc})$. The internal energy can also be
easily expressed in terms of classical quantities. Using the general
formula (\ref{Zsc}) we obtain
\begin{equation}
u_{sc}=-\frac{\partial}{\partial \beta}\ln {\cal
Z}_{sc}=\frac{1}{{\cal Z}_{sc}}\sum_{j}Z_{sc}^{(j)}\left\{{\cal
H}_0^{(j)}+\frac{1}{2}\frac{\partial}{\partial \beta}\ln({\rm
Tr}[M_0^{(j)}]-2) \right\}\; ,
\end{equation}
where we have used $\partial S/\partial \tau = -H=i {\cal H}$. Once
we get the Helmholtz and internal energies we can determine the
semiclassical expression for the entropy through relation
$s=(u-f)/T=k_B \beta(u-f)$. In the next section we give an example
of a system that classically violates the third law, and show that,
on the contrary, the semiclassical entropy does satisfy the
condition $s \rightarrow 0$ for $T \rightarrow 0$. Higher order
derivatives of thermodynamical interest can also be obtained
analytically, e. g., the specific heat $\partial u_{sc}/\partial T$:
\begin{equation}
c_{sc}=-k_B\beta^2 u_{sc}^2+\frac{k_B\beta^2}{{\cal
Z}_{sc}}\sum_{j}Z_{sc}^{(j)}\left[\left\{{\cal
H}_0^{(j)}+\frac{1}{2}\frac{\partial}{\partial \beta}\ln({\rm
Tr}[M_0^{(j)}]-2) \right\}^2-\frac{1}{2}\frac{\partial^2}{\partial
\beta^2}\ln({\rm Tr}[M_0^{(j)}]-2) \right]\;.
\end{equation}
In the next section we shall apply our formalism to simple examples
and compare the results with the pure classical and quantum ones.

\section{Applications}

\subsection{The simple harmonic Hamiltonian}
We first illustrate our semiclassical procedure with the simple
harmonic oscillator. The Hamiltonian is
\begin{equation}
{\cal H}=\frac{1}{2}p^2+\frac{1}{2}\omega^2q^2=\hslash \omega uv
\; \Rightarrow \; H=-i \hslash \omega uv\; . \label{hamh}
\end{equation}
The solutions of the equations of motion (\ref{hamilton3}) are
$u(\tau')=u'\,e^{-\omega \tau'}$, $v(\tau')=v''\,e^{\omega (\tau'
-\tau)}$, with $u(0)=z'$ and $v(\tau)=z'^*$. From these relations we
can write $u''=u'\,e^{-\omega \tau}$ and $v''=v'\,e^{\omega \tau}$,
or $\delta u''=\,e^{-\omega \tau} \delta u'$ and $\delta
v''=\,e^{\omega \tau} \delta v'$. Therefore
\begin{equation}
M_{uu}=\, e^{-\omega \tau} \qquad M_{vv}=\, e^{\omega \tau} \; .
\end{equation}
The only periodic trajectory is $u=v=0$, for which the complex
action vanish. Gathering the ingredients together we get the exact
quantum result for the partition function
\begin{equation}
{\cal Z}_{sc}=\frac{1}{\sqrt{e^{ \hslash \omega \beta}+e^{- \hslash
\omega \beta}-2}}=\frac{1}{2 \sinh\left(\frac{\hslash \omega
\beta}{2}\right)} \;.
\end{equation}
This expression leads to the well-known results for an ensemble of
non-interacting harmonic oscillators. In particular, we get for
$c_{sc}$ the Einstein expression for the specific heat of a
crystalline solid ($c_{sc}=c_{einstein}$).

\subsection{The quartic non-Euclidean Hamiltonian: quadratic approximation}

We now consider a more challenging system described by the
Hamiltonian
\begin{equation}
{\cal H}=\frac{1}{\hslash
\omega}\left(\frac{1}{2}p^2+\frac{1}{2}\omega^2q^2+ \lambda
\right)^2=\hslash \omega(uv+\alpha)^2 \; , \label{ham}
\end{equation}
with $\alpha = \lambda /\hslash \omega$. With the Wick rotation we
get
\begin{equation}
H=-i \hslash \omega(uv+\alpha)^2. \label{hamwick}
\end{equation}
Note that this Hamiltonian may be seen as a sum of a harmonic term
(proportional to $\alpha$) and an anharmonic quartic term
(independent of $\alpha$). Since $\alpha \sim \hslash^{-1}$, while
$uv \sim \hslash^{0}$, in the semiclassical limit ($\alpha >> 1$)
the quartic term must be seen as a perturbation added to the
harmonic Hamiltonian. Note also that the ${\cal H}$ can not be
written as $p^2/2+V(q)$. The equations of motion are
\begin{equation}
\dot{v}=2 \omega (uv+\alpha)v\;, \hspace{.2in} \mbox{and}
\hspace{.2in} \dot{u}=-2 \omega (uv+\alpha)u\; .
\end{equation}
Since $(uv+\alpha)=\sqrt{{\cal H}/\hslash \omega}$ is a constant
of the motion, the classical solution satisfying the boundary
conditions is
\begin{equation}
u(\tau')=u'\,e^{-\Omega \tau'}, \; \; \; v(\tau')=v''\,e^{\Omega
(\tau' -\tau)}\; \Rightarrow \; u''=u'\,e^{-\Omega \tau}, \; \; \;
v''=v'\,e^{\Omega \tau}\;,
\end{equation}
where $\Omega=2 \omega (u'v'+\alpha)$ and $\tau = \hslash \beta$. We
note that the only periodic orbit is the trivial one, although the
system does not involve a potential. The above solution is very
similar to that of the harmonic Hamiltonian but, in the present
case, the frequency $\Omega$ is energy dependent. The connection
between initial and final variations in a given trajectory can be
obtained from
\begin{equation}
\delta u''=\,e^{-\Omega \tau} \delta u'-\tau e^{-\Omega \tau}
\delta \Omega \;,\;\; \delta v''=\,e^{\Omega \tau} \delta v'+\tau
e^{\Omega \tau} \delta \Omega\; ,
\end{equation}
where $\delta \Omega =2 \omega (u'\delta v'+v'\delta u')$.
Therefore the relevant tangent matrix elements are
\begin{equation}
M_{uu}=(1-2 \omega \tau u' v')\, e^{-\Omega \tau}\;,
\end{equation}
and
\begin{equation}
M_{vv}=(1+2 \omega \tau u' v')\, e^{\Omega \tau}\; .
 \label{Mvv}
\end{equation}
On the trivial trajectory $u=v=0$ we obtain simply
\begin{equation}
M_{uu} = e^{-2 \alpha \hslash \omega \beta}, \quad M_{vv} = e^{2
\alpha \hslash \omega \beta}, \quad \frac{i}{\hslash}S_0 = -
\alpha^2 \hslash \omega \beta \; , \label{trivial}
\end{equation}
where we have set $\Omega = 2 \omega \alpha$. This leads to the
semiclassical result
\begin{equation}
{\cal Z}_{sc}=\frac{e^{-\alpha^2 \hslash \omega \beta}}{2 \sinh(
\alpha \hslash \omega \beta)}\;, \label{zscoh2}
\end{equation}
from which one can write
\begin{equation}
f_{sc}=\alpha^2 \hslash \omega +\frac{1}{\beta}\ln[2\sinh(\alpha
\hslash \omega \beta)]\; ,
\end{equation}
and
\begin{equation}
u_{sc}=\alpha^2 \hslash \omega +\hslash \omega \alpha \coth(\alpha
\hslash \omega \beta)\;.
\end{equation}
Not surprisingly, these expressions correspond to the exact quantum
result for the harmonic part of the Hamiltonian (\ref{ham}). In the
next subsection we develop a more accurate approximation. The exact
quantum partition function is given by ${\cal Z}=\sum_n e^{-\beta
E_n}$, where $\{E_n\}$ are the (non-degenerate) eigenvalues of the
quantum Hamiltonian $\hat{\cal H}=\hslash
\omega(\hat{a}^\dagger\hat{a}+\alpha)^2$. Thus
\begin{equation}
{\cal Z}=\sum_n e^{- \hslash \omega \beta (n+\alpha +1/2)^2 }\; .
\end{equation}
The classical function is given by
\begin{equation}
{\cal Z}_{class}=\int\frac{{\rm d}\,q{\rm d}p}{2\pi
\hslash}\,e^{-\beta {\cal H}}=\frac{\sqrt{ \pi}}{2 \,
\sqrt{\hslash \omega \beta}}\left[ 1-{\rm Erf}\left( \sqrt{\hslash
\omega \beta}\, \alpha\right)  \right]\; ,
\end{equation}
where Erf denotes the error function.

\subsection{The quartic non-Euclidean Hamiltonian: higher order approximation}

Higher order corrections to the semiclassical partition function can
also be obtained from (\ref{int}). We illustrate the method for the
Hamiltonian (\ref{ham}). Let us rewrite integral (\ref{int}) as
\begin{equation}
{\cal Z}_{sc}= \int \frac{du'\,dv''}{2\pi
i}A\;\exp\left\{\frac{i}{\hbar}\Phi(v'', u', \tau)_{v''=z'^*
,u'=z' }\;-z'^*z' \right\}\; , \label{int2}
\end{equation}
where $A=1/\sqrt{M_{vv}}$. Our procedure now is to expand the
argument of the exponential up to fourth order around the saddle
points, and the pre-factor up to second order (for more details
see appendix B of \cite{marcus}). We get
\begin{equation}
A \approx A_0+A_{(u)}\Delta u'+A_{(v)}\Delta
v''+\frac{1}{2}A_{(uu)}\Delta {u'}^2+\frac{1}{2}A_{(vv)}\Delta
{v''}^2+ A_{(uv)}\Delta u'\Delta {v''}\; ,
\end{equation}
and
\begin{equation}
\begin{array}{ll}
\frac{i}{\hbar}\Phi-z'^*z' &\approx
\frac{i}{\hslash}\Phi_0-u'_0v''_0+ \frac{i}{2\hslash}\Phi_{(uu)}
\Delta u'^2+\frac{i}{2\hslash}\Phi_{(vv)} \Delta v''^2
+\left(\frac{i}{\hslash} \Phi_{(uv)}-1 \right)\Delta u' \Delta v''\;
\\ \nonumber
&+\frac{i}{6 \hslash }\left[\Phi_{(uuu)} \Delta {u'}^3+3 \,
\Phi_{(uuv)}\Delta {u'}^2\Delta {v''}+3 \, \Phi_{(uvv)}\Delta
{u'}\Delta {v''}^2+\Phi_{(vvv)} \Delta {v''}^3\right]
\\ \nonumber
&+\frac{i}{24 \hslash}\left[\Phi_{(uuuu)} \Delta {u'}^4 + 4 \,
\Phi_{(uuuv)}\Delta {u'}^3\Delta {v''} +6 \, \Phi_{(uuvv)} \Delta
{u'}^2\Delta {v''}^2 \right.
\\ \nonumber
&\left. + 4 \, \Phi_{(uvvv)}\Delta {u'}\Delta {v''}^3+ \Phi_{(vvvv)}
\Delta {v''}^4  \right]\; ,
\end{array}
\end{equation}
where $A_{(u)}=(\partial A/ \partial u')_0$ and so on. For the
Hamiltonian (\ref{ham}) we find that the only non-vanishing terms
are the ones proportional to $A_{(uv)}$, $\Phi_{(uv)}$, and
$\Phi_{(uuvv)}$ besides the zeroth order terms (see appendix).
Therefore the partition function integral reads
\begin{equation}
\begin{array}{ll}
\tilde{\cal Z}_{sc}&= e^{\frac{i}{\hslash}\Phi_0} \int \frac{{\rm
d}(\Delta u')\,{\rm d}(\Delta v'')}{2\pi i} \left( A_0+
A_{(uv)}\Delta u'\Delta {v''}\right) \; \times \\
& \qquad \qquad \exp{\left \{ \left(\frac{i}{\hslash}\Phi_{(uv)}
-1\right) \Delta u' \Delta v'' +\frac{i}{4 \hslash}\,
\Phi_{(uuvv)}\Delta
{u'}^2\Delta {v''}^2\right \}}\;\\ \\
&\approx e^{- \omega \alpha^2 \hslash \beta} \int \frac{{\rm
d}(\Delta u')\,{\rm d}(\Delta v'')}{2\pi i} \left( A_0+
A_{(uv)}\Delta u'\Delta {v''}\right) \; \times \\
& \qquad \qquad \left(1 +\frac{i}{4 \hslash}\, \Phi_{(uuvv)}\Delta
{u'}^2\Delta {v''}^2\right ) \, \exp{\left \{
\left(\frac{i}{\hslash}\Phi_{(uv)} -1\right) \Delta u' \Delta
v''\right \}} \;, \label{int3}
\end{array}
\end{equation}
where we have assumed that the fourth order contribution in the
exponent is small if compared to the second order term, and the
tilde indicates the higher order approximation. The next step is
to diagonalize the quadratic form in the argument of the
exponential. This can be done with the variables $W$ and $K$ given
by
\begin{equation}
\Delta u'=\frac{1}{\sqrt{2 \, \gamma}}(K-iW)\;, \;\; \Delta
v''=-\frac{1}{\sqrt{2 \, \gamma} }(K+iW)\;,
\end{equation}
where $2\,\gamma \equiv (i\Phi_{(uv)}/\hslash -1)$. Integral
(\ref{int3}) becomes
\begin{equation}
\tilde{\cal Z}_{sc}=\frac{e^{- \omega \alpha^2 \hslash \beta}}{i
\, \gamma} \int \frac{{\rm d} W\,{\rm d} K}{2\pi i} \left[
A_0-\frac{A_{(uv)}}{2 \gamma}(W^2+K^2)+\frac{i A_0
\Phi_{(uuvv)}}{16\, \hslash \gamma^2}(W^2+K^2)^2\right]\,
e^{-W^2-K^2} \; ,
\end{equation}
where we have excluded terms of order higher than four. The
resulting partition function is
\begin{equation}
\tilde{\cal Z}_{sc}={\cal Z}_{sc}\left( 1-\frac{A_{(uv)}}{2 \gamma
A_0} -\frac{\Phi_{(uuvv)}}{i8 \hslash \gamma^2 }\right)\;,
\end{equation}
where ${\cal Z}_{sc}$ is given by expression (\ref{zscoh2}). We have
find that (see appendix)
\begin{equation}
A_0=e^{-\alpha \hslash \omega \beta}\;, \;\; A_{(uv)}=-2 \hslash
\omega \beta \,e^{-3\,\alpha \hslash \omega \beta}\;,\;\;
\Phi_{(uuvv)}=4i\, \hslash^2 \omega \beta \,e^{-4\,\alpha \hslash
\omega \beta}\;, \label{terms}
\end{equation}
and, therefore,
\begin{equation}
\tilde{\cal Z}_{sc}=\frac{e^{- \alpha^2 \hslash \omega \beta}}{2
\sinh( \alpha \hslash \omega \beta)}\left( 1- \frac{\hslash \omega
\beta}{2} \left[ \sinh(\alpha \hslash \omega
\beta)\right]^{-2}\right)\;.
\end{equation}

\begin{figure}
  \includegraphics[width=5.7cm,angle=-90]{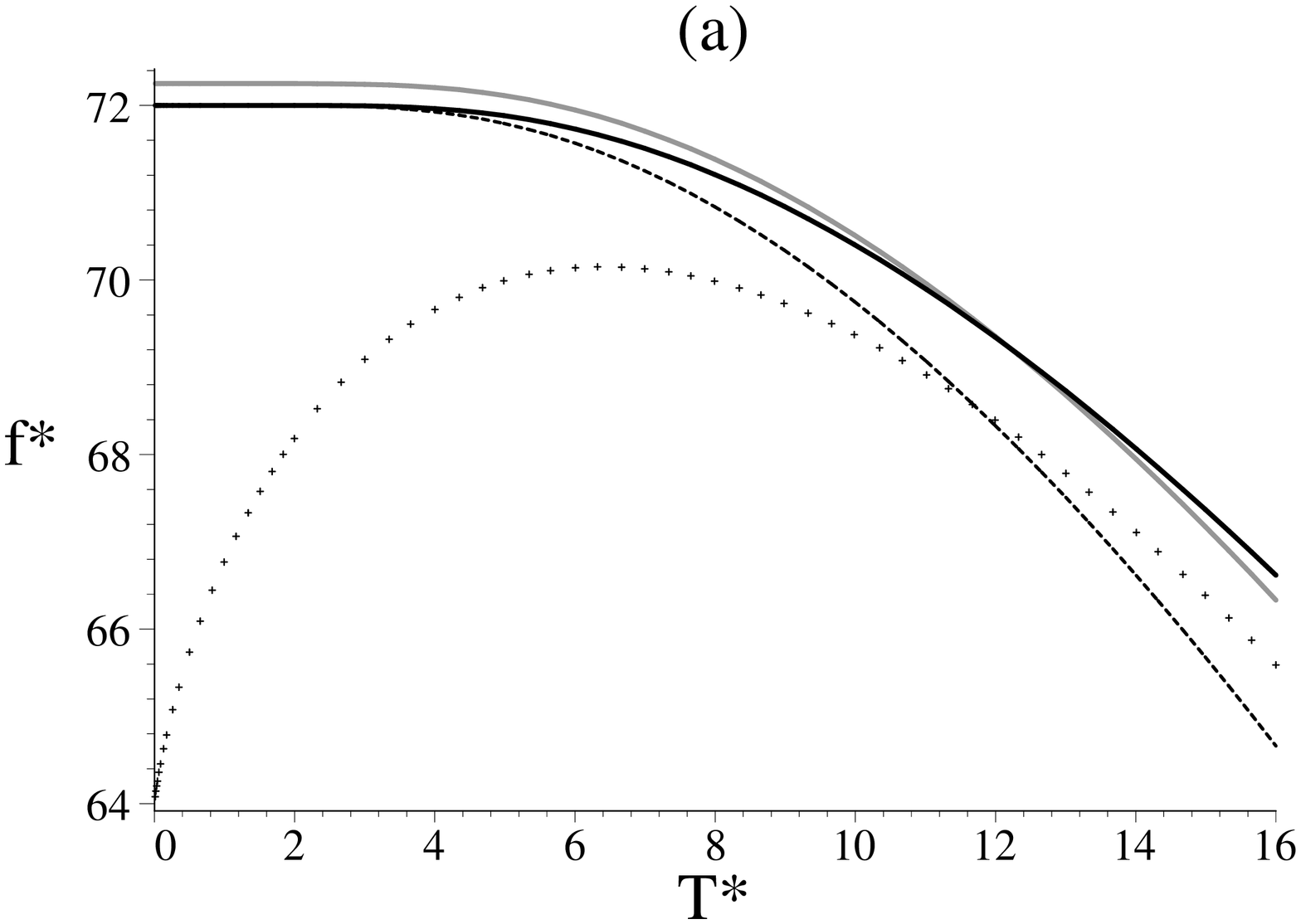}
  \includegraphics[width=5.7cm,angle=-90]{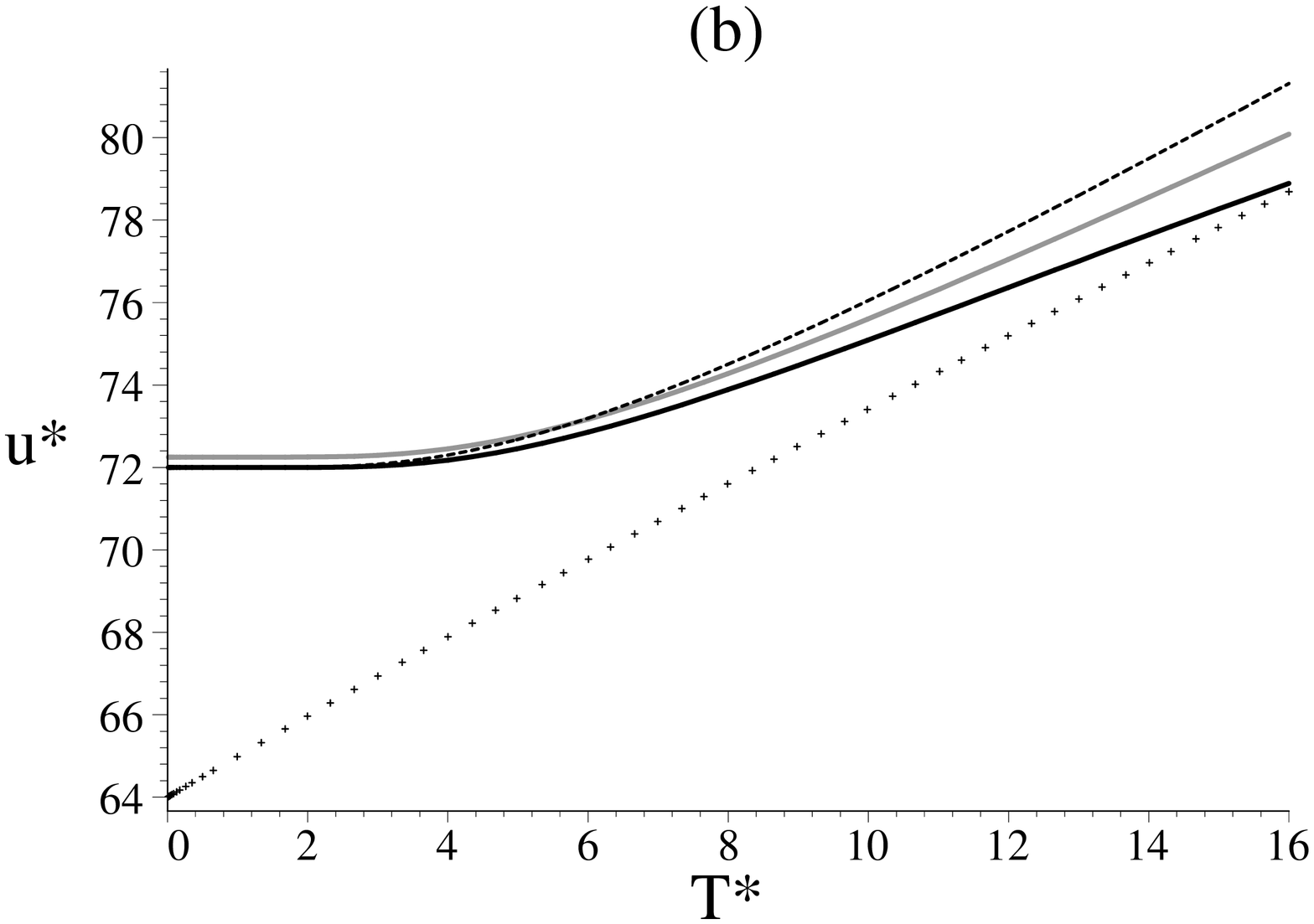}
  \caption{Scaled Helmholtz free
  energy (a) and scaled internal energy (b): higher order semiclassical (black),
  quantum (gray), classical (crosses), and quadratic semiclassical (dashed line).}
  \label{fig1}
\end{figure}

\begin{figure}
  \includegraphics[width=7.5cm,angle=-90]{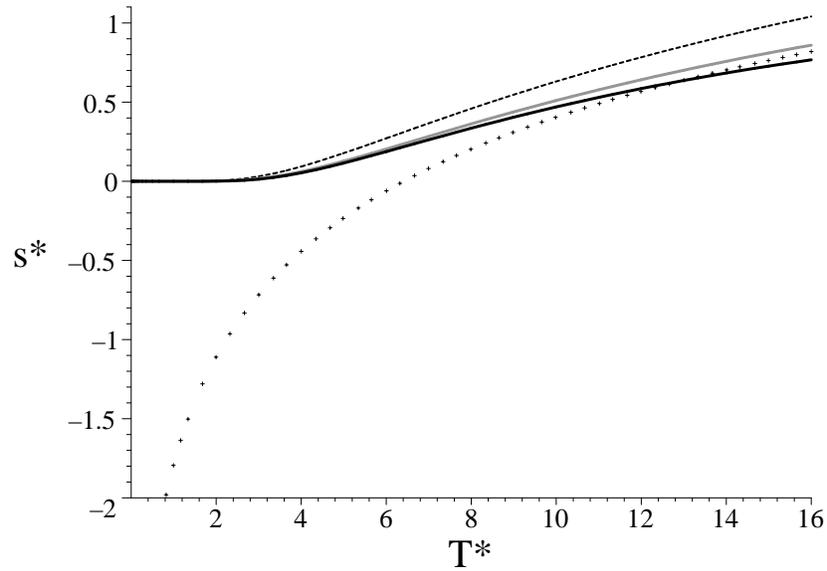}
  \caption{Scaled entropy: higher order semiclassical (black), quantum (gray),
   classical (crosses), and quadratic semiclassical (dashed line). The third law of
   thermodynamics is satisfied by the semiclassical curves.}
  \label{fig2}
\end{figure}

In figure \ref{fig1} we compare the scaled thermodynamic potentials
$f^*=f/\hslash \omega$ and $u^*=u/\hslash \omega$ as functions of
the reduced temperature $T^{*} =k_B T/\hslash \omega$, with
$\alpha=8.0$, for the four cases: higher order semiclassical
(black), quantum (gray), classical (crosses), and quadratic
semiclassical (dashed line). As expected the classical results tend to
the exact quantum ones for high temperatures, while they completely
fail to describe the physical behavior at $T \rightarrow 0$. In
contrast, the semiclassical approximations are very precise in the
low temperature limit. In particular we note that the semiclassical
entropy satisfies the third law of thermodynamics while the
classical function does not. This is shown in figure \ref{fig2},
where the scaled entropy $s^*=s/k_B$ is plotted as a function of
$T^*$.

\subsection{Two-Well Potential}

\begin{figure}
  \includegraphics[width=5.7cm,angle=-90]{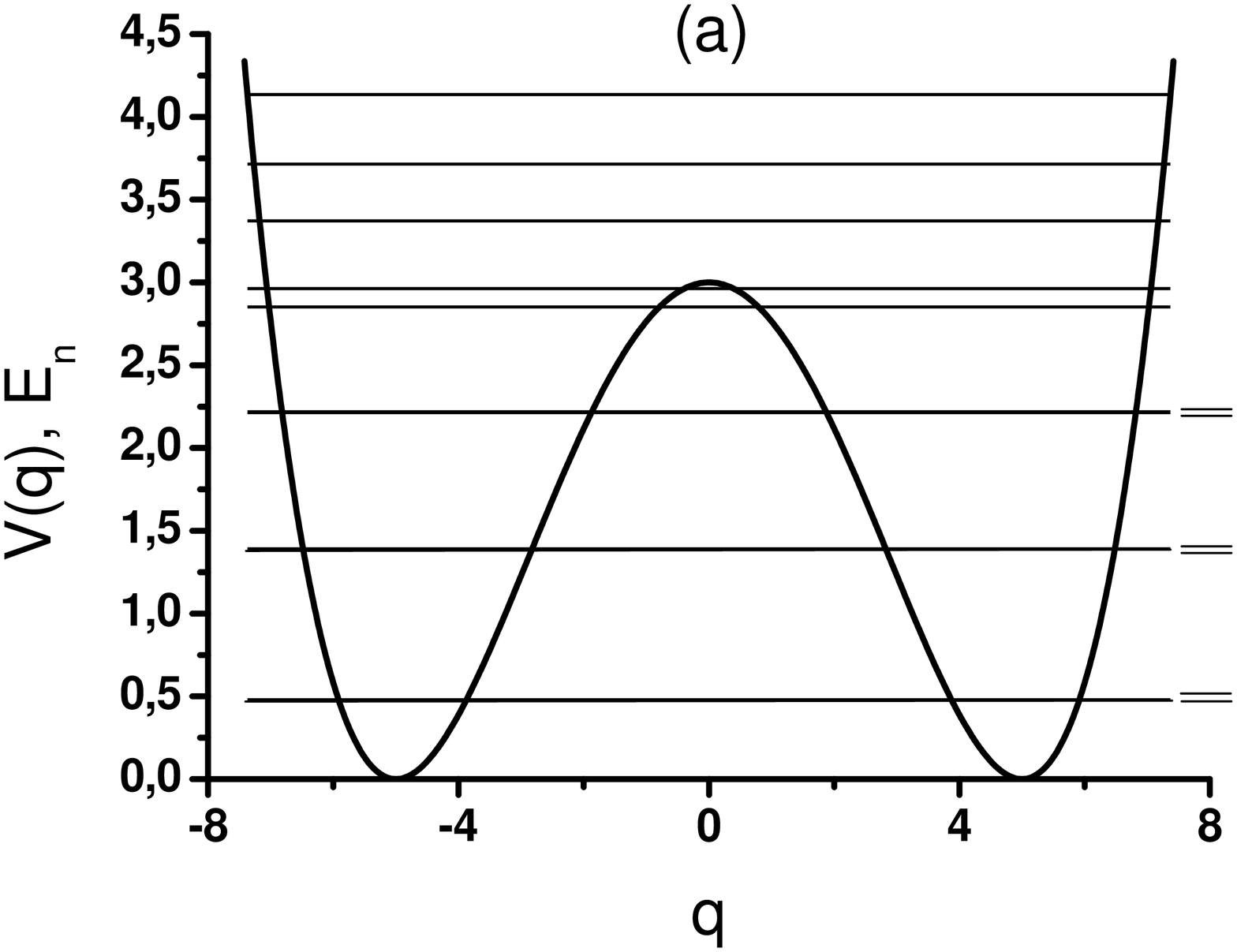}
  \includegraphics[width=5.7cm,angle=-90]{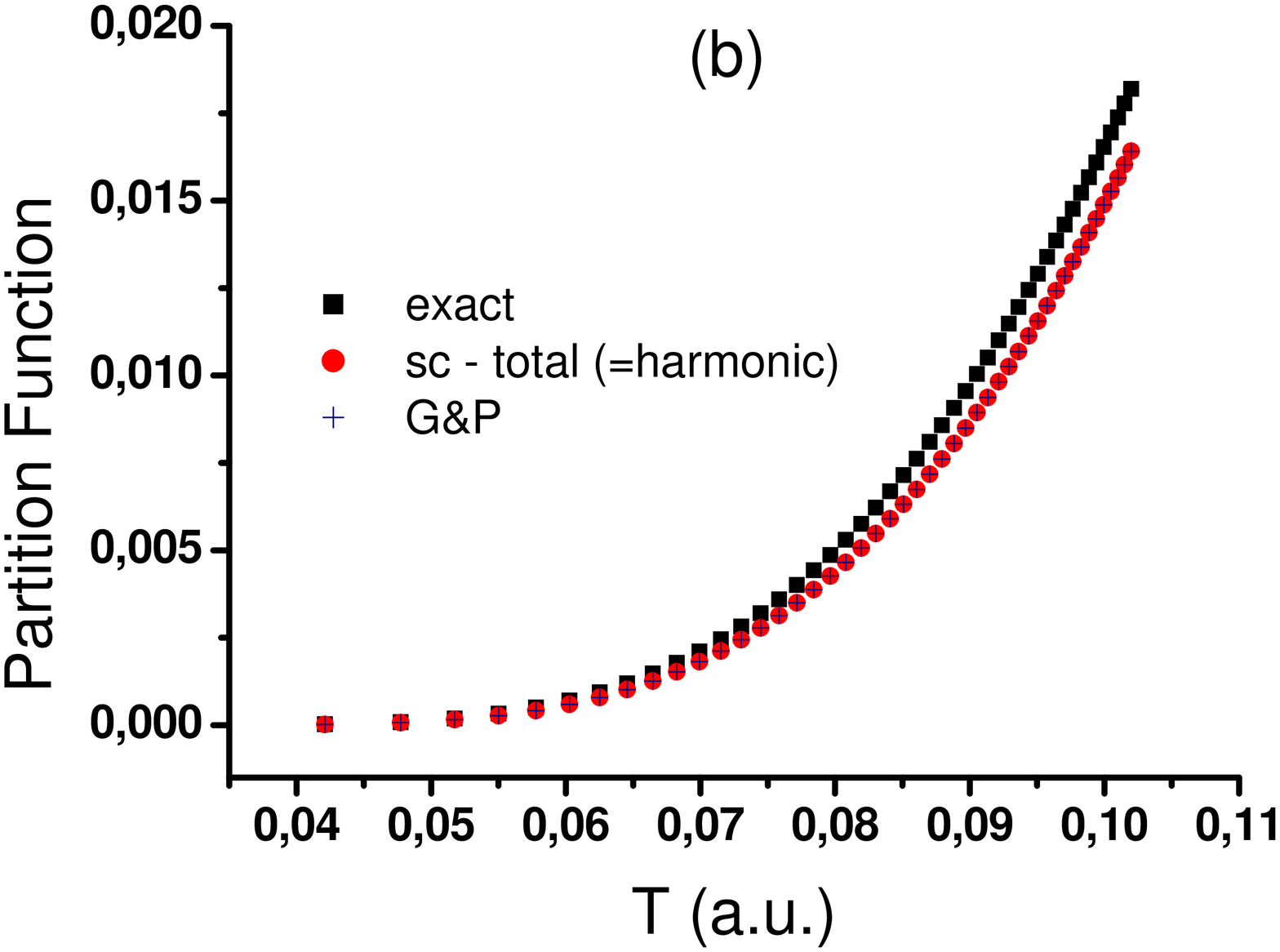}
  \caption{The deep double-well is shown together with the first eleven energy
   levels in (a). In (b) the exact, semiclassical, and G\&P results are
   compared.}
  \label{fig3}
\end{figure}

\begin{figure}
  \includegraphics[width=5.7cm,angle=-90]{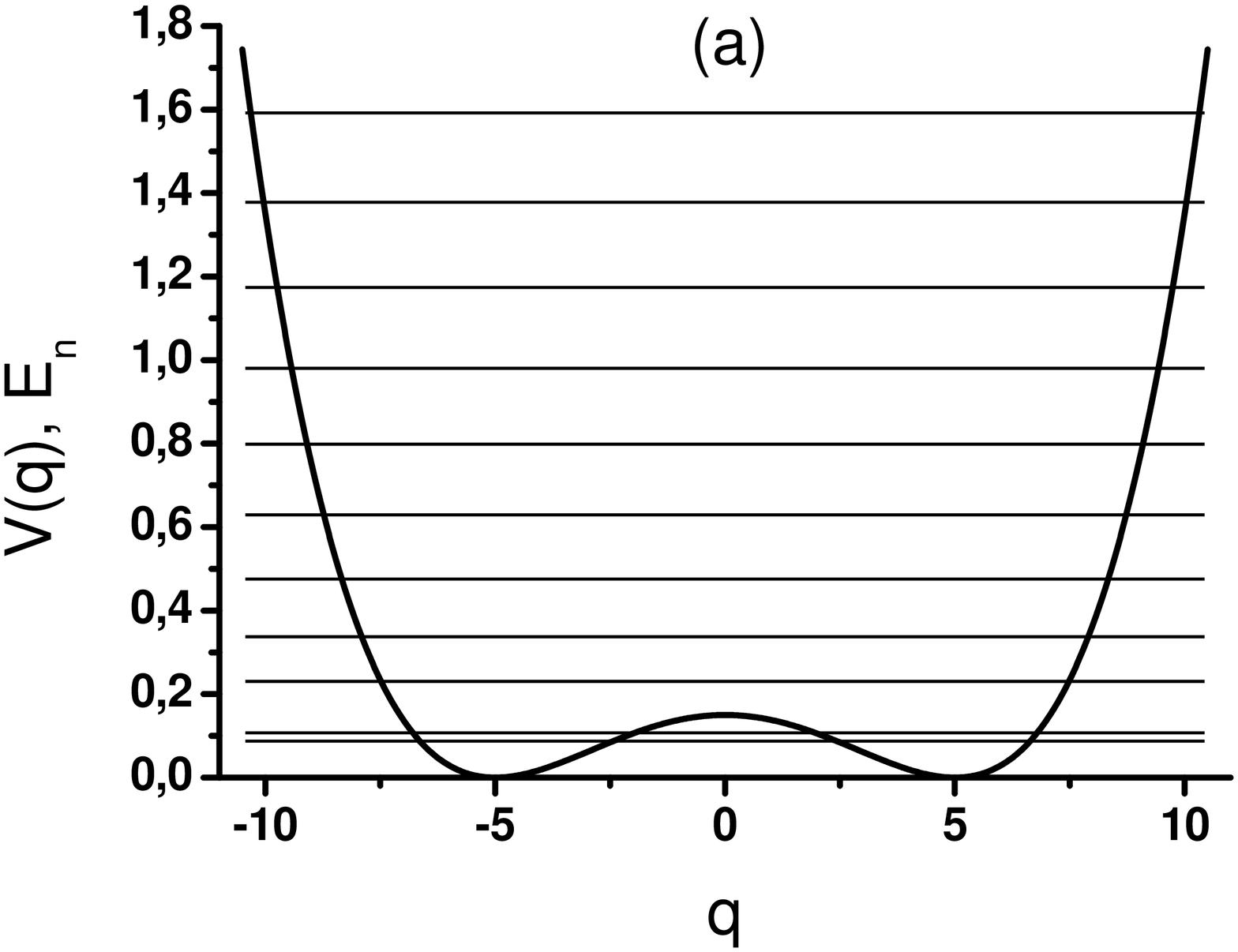}
  \includegraphics[width=5.7cm,angle=-90]{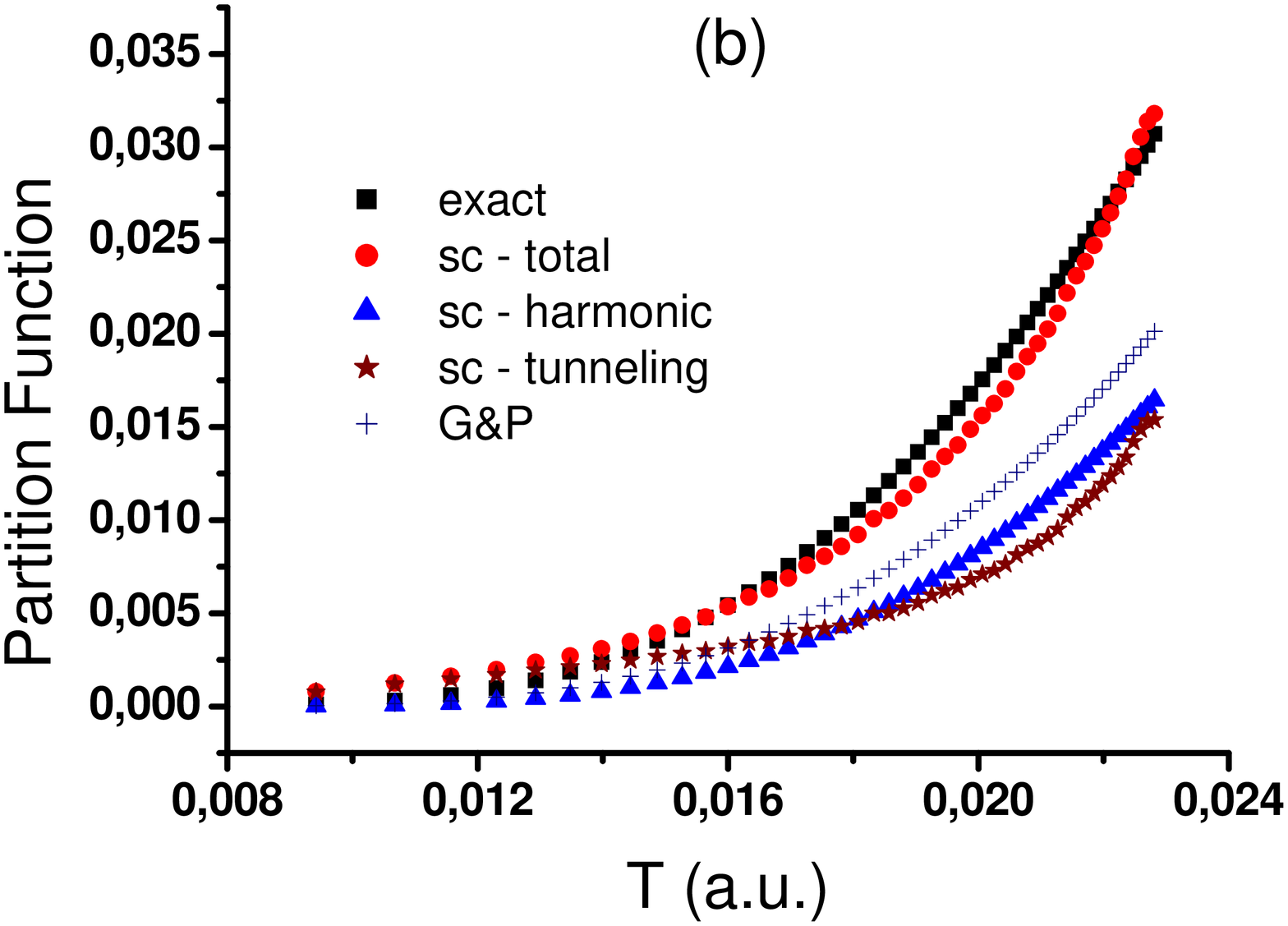}
  \caption{The shallow double-well is shown together with the first eleven energy levels
   in (a). In (b) the exact, total semiclassical, harmonic contribution, tunneling contribution,
   and G\&P results are displayed.}
  \label{fig4}
\end{figure}

Here we address a system with a quartic potential function $V =
A\,q^4 -B\,q^2+C$, with $A,B>0$. If we denote the height of the
local maximum located at $q=0$ by $\Delta E$ and the distance
between the two symmetric minima by $2a$, the potential is written
as
\begin{equation}
V(q)=\Delta
E\left[\left(\frac{q}{a}\right)^4-2\left(\frac{q}{a}\right)^2
+1\right] \;, \label{2min}
\end{equation}
such that $V(\pm a)=0$. In this case we apply equation (\ref{Zmm})
directly. The harmonic contribution is $2 \times 1/[2\sinh(\hslash
\sqrt{2\Delta E}\beta/a)]$, while the tunneling term must be
determined numerically. There are two distinct regimes to be
considered. If $\Delta E$ is large enough to encompass several
energy levels, the harmonic contribution should be dominant over the
tunneling one, since the two minima hardly `see' each other. In this
case the trajectories connecting the wells have large actions,
giving exponentially small contributions to the semiclassical
partition function. In contrast, when $\Delta E$ is comparable with
the energy of the ground state, the tunneling term should give a
relevant contribution. These statements are verified in the
numerical computation of ${\cal Z}_{sc}$ for $\Delta E = 3.0$ and
$\Delta E = 0.15$ with $a=5$ in both cases (we use arbitrary units
for which $\hslash = 1$ and $k_B=1$). In the first case (deep well)
the ground state has an approximate energy of $0.5$, and there are
eight levels (four doublets) with energy smaller than $\Delta E$, as
shown in fig. \ref{fig3}(a). Figure \ref{fig3}(b) shows the quantum
partition function and ${\cal Z}_{sc}$, where we have used only the
harmonic term, since the tunneling contribution is about three
orders of magnitude smaller. The approximation derived by Gildener
and Patrascioiu \cite{GP} (see also \cite{Har78}) specifically for
the potential (\ref{2min}) is also displayed. This approximation and
the semiclassical one produce almost identical results in this case.
In the case of the shallow well there are only two levels with
energy below $\Delta E=0.15$ (see fig \ref{fig4}(a)). As expected
the harmonic contribution is unable to give a good approximation to
${\cal Z}$ and the tunneling term must be considered. The overall
semiclassical result is in very good agreement with the exact one,
as can be verified from fig \ref{fig4}(b). In this case, the G$\&$P
formula does not agree with the quantum result. In the computation
of the tunneling term we have not considered multi-period
trajectories, since we have found that the corresponding
contributions are vanishingly small.

\section{Spin Systems}

As discussed in the beginning of section II.c, the coherent state 
propagator can be described by a variety of semiclassical expressions.
For simplicity we chose the one involving the classical Hamiltonian
${\cal H}$. For spin systems there is no direct classical Hamiltonian and
we have to use a different representation. Our starting point 
is the spin coherent state, defined by $|w \rangle=(1+w^*w)^{-s}
\exp\{ w\hat{S}_+/ \hslash\}|s,-s \rangle$, where $\hslash s$ is the
total spin and $w$ is a complex number. The Q-representation of the
Hamiltonian operator is defined as ${\cal H} \equiv \langle w| 
\hat{\cal H}|w \rangle$. This function appears naturally in the first
form of path integral suggested by Klauder and Skagerstan
\cite{Klau85}, whose semiclassical limit requires the addition of an
extra term in  (\ref{action1}) \cite{marcus}, the so-called
Solari-Kochetov phase  \cite{solari,kochetov}:
\begin{equation}
{\cal I}_{SK}=\frac{1}{4}\int_0^t\left[ \frac{\partial}{\partial w^*}
\left(\frac{(1+w^*w)^2}{2s}\frac{\partial {\cal H}}{\partial w} \right)+ 
\frac{\partial}{\partial w}\left(
\frac{(1+w^*w)^2}{2s}\frac{\partial {\cal H}}{\partial w^*} \right)
\right]{\rm d}t'\; .
\end{equation}
Furthermore, since $w$ and $w^*$ are not canonically conjugated, it is
convenient to define new canoncial variables by \cite{marcus2}
\begin{equation}
Q=i\sqrt{\frac{s}{1+w^*w}}\,(w^*-w)\;, \;\;
P=\sqrt{\frac{s}{1+w^*w}}\,(w^*+w)\;.
\end{equation}
This transformation is the classical analogue of the Holtein-Primakoff
bosonization procedure \cite{hp}. In these variables, the equations
satisfied by the stationary phase trajectories that contribute to the
semiclassical propagator are the usual Hamilton's equation of motion
\begin{equation} 
\frac{{\rm d} Q}{{\rm d} t}=\frac{\partial {\cal H}}{\partial P}\; ,
\;\; \frac{{\rm d} P}{{\rm d} t}=-\frac{\partial {\cal H}}{\partial
Q}.   
\end{equation}

We remark that the Solari-Kochetov term does not change the stationary
exponent condition since it already involves second derivatives of the 
Hamiltonian. With this procedure the formalism we have developed can be 
readily applied to spin systems. The semiclassical partition function reads
\begin{equation}
{\cal Z}_{sc}=\sum_j\frac{e^{\frac{i}{\hslash} (S_0^{(j)}
+{\cal I}_{SK}^{(j)})}}{\sqrt{{\rm
Tr}[M^{(j)}_0]-2}}\; , \;\; \mbox{or} \;\;
{\cal Z}_{sc}=\sum_j\frac{e^{-\frac{1}{\hslash} (\overline{S}_0^{(j)}
+\overline{\cal I}_{SK}^{(j)})}}{\sqrt{{\rm
Tr}[M^{(j)}_0]-2}}\; , 
\label{Zscspin}
\end{equation}
for transformations (\ref{hamiltons}) and (\ref{newton}), respectively.

As a simple example we consider a spin in a uniform magnetic
field. The Hamiltonian operator is $\hat{\cal H}=\omega\hat{S}_z$,
leading to the exact partition function 
\begin{equation}
{\cal Z}=e^{\hslash \omega \beta s} \sum_{n=0}^{2s} e^{-\hslash \omega \beta n}
=\frac{e^{\hslash \omega \beta (s+1/2)}-e^{-\hslash \omega \beta (s+1/2)}}
{e^{\hslash \omega \beta /2}-e^{-\hslash \omega \beta /2}}\;.
\end{equation}
From the definition of spin coherent states we obtain
\begin{equation}
{\cal H}=-\hslash \omega s \left( \frac{1-w^*w}{1-w^*w}\right)=
\hslash \omega \left( \frac{P^2}{2}+\frac{Q^2}{2}\right)-\hslash \omega s\;,
\label{hspin}
\end{equation}
from which one sees that the only contributing orbit is the trivial one.
This yields $iS/\hslash=\hslash \omega \beta s$ and 
$i{\cal I}_{SK}/\hslash=\hslash \omega \beta/2$.
The corresponding semiclassical partition function is
\begin{equation}
{\cal Z}_{sc}=\frac{e^{\hslash \omega \beta (s+1/2)}}{e^{\hslash \omega 
\beta /2}-e^{-\hslash \omega \beta /2}}\;,
\end{equation}
which differs from the exact result by a factor that is exponentially
small in the low temperature and large spin limits. At first glance it
might be expected that the semiclassical result should be exact, since
the effective Hamiltonian is quadratic. However, it is important to
note that the classical problem we are dealing with is  not completely
equivalent to that of a harmonic oscillator. In the present case,  
the phase-space is compact since ${\cal H} \le \hslash \omega s$ (
see Eq. (\ref{hspin})). The two problems are strictly equivalent only if 
$s \rightarrow \infty$, when the semiclassical result is exact. 

\section{Concluding Remarks}

We have obtained a semiclassical trace formula for the canonical
partition function in the low temperature limit that does not require
sampling and numerical integration in phase-space. The examples
presented in sections III and IV show that, in spite of the failure of
the classical partition function in describing thermal effects at low
temperatures, the appropriate combination of classical ingredients
appearing in the semiclassical formula is indeed capable of giving
very accurate results in this limit. The quantum corrections come
precisely in the way we combine the classical functions. The two and
three-dimensional extensions of the formalism are straightforward in
the case of single minimum potentials.  

In the case of multiple minima, the calculation of the non-trivial
periodic orbits is more involved due to the higher
dimensionality. However, contrary to the situation in usual trace
formulas, only very long orbits should contribute in the low
temperature regime. In particular, heteroclinic orbits connecting the
top of the inverted wells should be of great importance. Calculations
for two-dimensional potentials are currently under way. Other
potentially interesting perspectives are also open, such as the 
inclusion of spin-orbit interactions, whose semiclassical propagator
has been recently derived \cite{piza}.

\acknowledgments This work was partially supported by FAPESP
(04/13525-5 and 03/12097-7) and CNPq.
\begin{appendix}
\section{ Expansion coefficients }
In this appendix we indicate how to obtain the equalities
(\ref{terms}). The first expression is immediate
\begin{equation}
A_0=\sqrt{\frac{1}{M_{vv}}}=e^{-\omega \alpha \hslash \beta}\;,
\end{equation}
where we have used the second relation in (\ref{trivial}). Now let
us calculate $A_{(u)}$. We have
\begin{equation}
A_{(u)}=\left(\frac{\partial A}{\partial u'}
\right)_0=-\left(\frac{1}{2\,M_{vv}^{3/2}}\frac{\partial
M_{vv}}{\partial u'}\right)_0\;.
\end{equation}
From equation (\ref{Mvv}) we have $M_{vv}=(1+2 \omega \tau u'
v')\, e^{\Omega \tau}$, which in terms of the independent
variables $u'$ and $v''$ becomes $M_{vv}=(e^{\Omega \tau}+2 \omega
\tau u' v'')$. Therefore
\begin{equation}
\frac{\partial M_{vv}}{\partial u'}=2\omega \tau v''+\tau
\frac{\partial \Omega}{\partial u'}\, e^{\Omega \tau} \;.
\end{equation}
One can implicitly determine the derivative of $\Omega$ starting
from $\Omega=2 \omega (u'v'+\alpha)=2 \omega (u'v''\,e^{-\Omega
\tau}+\alpha)$. The result is
\begin{equation}
\frac{\partial \Omega}{\partial u'}=\frac{2\omega v''\,e^{-\Omega
\tau}}{1+2\omega\tau u'v''\,e^{-\Omega \tau}} \; \Rightarrow \;
\left(\frac{\partial \Omega}{\partial u'}\right)_0 =0\;.
\label{domega}
\end{equation}
Consequently we have $\left(\frac{\partial M_{vv}}{\partial
u'}\right)_0=0$ and $A_{(u)}=0$. A similar calculation shows that
$A_{(v)}=0$. For $A_{(uv)}$ we get
\begin{equation}
A_{(uv)}=-\frac{1}{2}\left( \frac{1}{M_{vv}^{3/2}}\frac{\partial^2
M_{vv}}{\partial u' \partial v''}-
\frac{3}{2M_{vv}^{5/2}}\frac{\partial M_{vv}}{\partial
u'}\frac{\partial M_{vv}}{\partial
v''}\right)_0=-\frac{1}{2}\left(
\frac{1}{M_{vv}^{3/2}}\frac{\partial^2 M_{vv}}{\partial u'
\partial v''}\right)_0\;,
\end{equation}
with
\begin{equation}
\left(\frac{\partial^2 M_{vv}}{\partial u' \partial
v''}\right)_0=\tau\left(\frac{\partial^2 \Omega}{\partial u'
\partial v''}\right)_0 \, e^{2 \omega \alpha \tau}+2\omega \tau\;.
\end{equation}
From equation (\ref{domega}) we can show that
\begin{equation}
\left(\frac{\partial^2 \Omega}{\partial u' \partial
v''}\right)_0=2\omega \,e^{-2\omega \alpha \tau} \; \Rightarrow \;
\left(\frac{\partial^2 M_{vv}}{\partial u' \partial
v''}\right)_0=4\omega \tau \;,
\end{equation}
which leads to $A_{(uv)}=-2\omega \tau \, e^{-3\omega \alpha
\tau}$. Using the same strategy one can prove that
$A_{(uu)}=A_{(vv)}=0$, that all third order derivatives of $\Phi$
vanish at the trivial orbit, and that the only non-vanishing
fourth order derivative is
\begin{equation}
\Phi_{(uuvv)}=\left( \frac{i \hslash}{M_{vv}^{2}}\frac{\partial^2
M_{vv}}{\partial u' \partial v''} \right)_0=4 i \hslash\ \omega
\tau \, e^{-4\omega \alpha \tau}\;.
\end{equation}
\end{appendix}


%
\newpage
%
\end{document}